\documentclass[prb,showpacs,preprintnumbers,amsmath,amssymb]{revtex4}

\usepackage{graphicx}
\usepackage{dcolumn}
\usepackage{bm}

\begin{document}

\title{Cubic magnets with Dzyaloshinskii-Moriya interaction at low T}

 \author{S. V. Maleyev}
\affiliation{Petersburg Nuclear Physics Institute, Gatchina, St.\ Petersburg 188300, Russia}
\email{maleyev@sm8283.spb.edu}
\newcommand{\te}{\textbf}
\newcommand{\m}{\mathbf}
\newcommand{\ks}{\hat\xi_\m R}
\newcommand{\et}{\hat\eta_\m R}

\newcommand{\ze}{\hat\zeta_\m R}

\begin{abstract}
  Ground state and spin-wave spectrum of cubic magnets with the Dzyaloshinskii-Moriya interaction (DMI) such as 
$M n S i$  and   $F e G e $ are studied theoretically. Following interactions are taken into account: conventional isotropic exchange, the DMI, anisotropic exchange, magnetic dipole interaction and Zeeman energy. In the classical approximation these interactions determine the helix wave -vector $\m k$ and critical field $H_c$  for  the transition to the "ferromagnetic" spin configuration. This field depends on the sample form due to the demagnetization. The linear spin-wave theory is developed. The spin-wave spectrum depends strongly on the magnetic field. At $H>H_c$ we have quadratic spectrum  with the gap linearly increasing with the field. Below $H_c$ the spectrum is gapless and strongly anisotropic.  It is a result of incommensurate magnetic structure when the DMI  breaks the total spin conservation law and  umklapp processes appear connecting the spin-wave excitations  with momenta $\m q$ and $\m{q\pm k}$ with different energies. For $\m q$ along $\m k$ the spectrum is linear. For other $\m q$ directions it have very complex form  determined by solution of  infinite set of linear equations connecting the states with $\m q$ and $\m q\pm n\m k$ where $n=1,2,...$. Restriction to $n=1$ gives six equations which general solution remains  complex.  For $\m{q\perp k}$ there are two modes: one has  the gap equal to $A k^2\sqrt{2} $ where  $A $ is the spin-wave stiffness  at  $q\gg k$.  The second is  gapless and proportional to $\m q^2_\perp$. At $q_\perp\gg k$ all branches merge and the anisotropy of the spectrum disappears. These results changes insignificantly in $n=2$ approximation.

The classical energy depends on the field component along the helix axis $\m k$ only. However it is known experimentally that rather weak perpendicular field  rotates the helix  and its axis is settled  along the field. This quantum phenomenon is a consequence  of the umklapps too.  The spin-wave spectrum  is unstable at infinitesimal perpendicular field. If the gap $\Delta$ is introduced   the spectrum becomes stable if $g\mu_B H_\perp<\Delta$ In this field  there are the magnetization along $\m H_\perp$ and deformation of the helix. The gap appears due to cubic anisotropy and the spin-wave interaction  considered in the Hartree-Fock approximation.

Peculiar  properties   of the ESR and neutron scattering in the helical magnets are considered and possibilities of corresponding experimental studies are discussed.
\end{abstract}


\maketitle
\section{Introduction}
Itinerant cubic magnets $M n S i$, $F e G e$, $F e S i C o$,  etc have attracted a lot of attention last years due to their specific electronic and magnetic properties. The former are characterized by closeness to quantum phase  transition, which is achieved at high pressure (see \cite{yu}, \cite{CP}  and references therein). Corresponding theory was developed in \cite{kir}. The last is related to the  $P2_13$ symmetry which allows the Dzyaloshinskii-Moriya interaction (DMI) responsible for magnetic helix structure. In cubic crystals the DMI  fixes the  the sense of the helix (right or left-handed spiral) but can't determine its direction. It is stabilized by very weak anisotropic exchange interaction (AEI) only \cite{nak} and \cite{bak}. The helix structure is very sensitive to external magnetic field. Two field-induced transitions were observed \cite{leb}, \cite{koy} and \cite{ok} and references therein. In low field it is transition to the state with the helix axis along the field. 
Then, there is the second  transition to the "ferromagnetic" state. These transitions were described by the mean-field theory \cite{plum} which contains a lot of phenomenological parameters. It should be  noted also  that critical properties of $M n S i$ near transition temperature $T_c$ are very unusual too. Corresponding experimental studies and its theoretical explanation  may be found in Refs. \cite {ros} \cite{geo} and \cite{gr}.

In this paper we present theoretical description of the low-temperature properties of the cubic magnets with the DMI. We microscopically evaluate the ground-state energy and spin-wave spectrum. In particular we demonstrate that the critical fields mentioned above are related to the parameters of the spin-wave spectrum in agreement with existing experimental data.

The outline of this paper is follows. In Sec. II the theoretical model is formulated. Along with the exchange interaction , the DMI and AEI considered in \cite{bak} we include also the magnetic dipole interaction and  the Zeeman energy. Following  \cite{bar} we introduce  all three spin components  in each lattice point of the helix and demonstrate that in "ferromagnetic"  state the perpendicular spin components responsible for the spin-waves remain rotating. Sec. III is devoted to consideration of the ground state energy. The critical field for the transition to the "ferromagnetic " state is determined with the demagnetization corrections. In Sec. IV the spin-wave Hamiltonian is considered. It consists of two parts: direct and umklapp. The first part is a conventional spin-wave Hamiltonian. Its diagonalization determines the spin-wave spectrum. The second one describes transitions between spin-waves with momenta $\m q$ and $\m{q\pm k}$ and  different energies. These transitions are a result of  the incommensurate helical structure and low symmetry of the DMI. This umklapp interaction is considered in Sec.V. It is shown that the umklapps created by the field perpendicular to the helix wave-vector $\m k$ are  a cause  of its turn to the direction of the field. At zero field the umklapps bring to strong anisotropy of the spin-wave spectrum: excitations with momentum along and perpendicular to $\m k$ have different energies. Spin configuration in weak perpendicular field is discussed in Sec. VI. Origin of  the spin-wave gap $\Delta$ is studied in Sec. VII. There are two contributions to $\Delta^2$: the spin-wave interaction considered in the Hartree-Fock approximation and the cubic anisotropy. If the last contribution is negative and  sufficiently strong instead of  the helical magnetic order the chiral spin liquid state is realized.  Possibilities  of the ESR and neutron scattering studies of  the dynamical phenomena investigated  in the paper is discussed in Sec. VIII. The principal results  are summarized and  discussed in Sec. IX. Some details of calculations are presented in Appendixes A--D.  
 
\section{model}
There are following principal interactions in  cubic  magnets without center of inversion: i) conventional exchange interaction, ii) the DMI, iii)the AEI, iv)magnetic dipole interaction and Zeeman energy.   We investigate  spin  configuration in magnetic  field and the spin-wave spectrum  using these  interactions. Then we discuss  role of the  cubic anisotropy in formation of the spin-wave gap.

 There are four magnetic  ions in the unit sell of the magnets under study. We will consider here the total spin of the unit sell as a basic magnetic entity, completely neglecting   internal movements of the cell spins and corresponding optical spin-wave branches. Interested in spin dynamics, we can use the magnetic density approximation \cite{pit} or the effective spin-lattice model with total spin of the unit cell $\m S(\m R)$ and usual commutation relations 
\begin{equation}
	[S_\alpha(\m R),S_\beta(\m R)]=i\epsilon_{\alpha\beta\gamma}S_\gamma(\m R)
\end{equation}
The DMI destroys  ferromagnetism and is responsible for the long periodical spin-density wave \cite{dz}, \cite{nak},\cite{bak}. For each spin we have  right-handed local orthogonal coordinate frame with basic unit vectors $\ze$, $\et$ and $\ks$ which for  helical structure are given by \cite{bar}
\begin{equation}
\begin{aligned}
	\ze &=\hat c\sin \alpha+(\hat a\cos\m{k\cdot R}+\hat b \sin\m{k\cdot R})\cos\alpha,\\
	\et&=-\hat a\sin\m{k R}+\hat b \cos \m{k R},\\
	\ks &=\hat c\cos \alpha-(\hat a \cos\m{k\cdot R}+\hat b\sin\m{k\cdot R})\sin\alpha.
\end{aligned}	
\end{equation}
where  $\hat a\times \hat b=\hat c$, $\hat b\times \hat c=\hat a$ and $\hat c\times \hat a=\hat b$.\\
Corresponding projections of the spin operators have well known form
\begin{equation}
\begin{aligned}
S_\m R ^\zeta &=S-(a^+a)_\m R,\\
S^\eta_\m R&=\frac{1}{i}\sqrt{\frac{S}{2}}\left[a_\m R-a^+_\m R-\frac{(a^+a^2)_\m R}{2S}\right],\\
S^\xi_\m R&=\sqrt{\frac{S}{2}}\left[a_\m R+a^+_\m R-\frac{(a^+a^2)_\m R}{2S}\right],
\end{aligned}
\end{equation}
here $S$ is a parameter connected to the cell magnetization by  $M=g\mu_B S/a^3$ where   $\mu_B>0$ and  $g\simeq 2$.
For $Mn Si$  magnetic moment per spin $0.4\mu_B$ is strongly reduced in comparison with the paramagnetic state value $1.4\mu_B$ and the unit-cell spin $S=0.2\times 4=0.8$  \cite{koy} .

For the above-mentioned principal interactions in the $\m R$ space we have
\begin{equation}
\begin{aligned}
H&=H_{EX}+H_{DM}+H_{AE}+H_Z H_D\\
H_{EX}&=-\frac{1}{2}\sum J_{\m {R R'}} \m S_\m R\cdot \m S_\m {R'},\\
H_{DM}&=\frac{1}{2}\sum D_{\m{R R'}}( \nabla-\nabla' )\m{S_R\times S_{R'}},\\
H_{AE}&=\frac{1}{2}\sum F_{\m R\m R'}[(\nabla_x S^x_\m R)(\nabla_x S^x_{\m R'}),\\&
+(\nabla_y S^y_\m R)(\nabla_y S^y_{\m R'})+(\nabla_z S^z_\m R)(\nabla_z S^z_{\m R'})],\\
H_D&=\frac{g\mu_B}{2}\sum [\m S_{\m R}\cdot \m S_{\m R}-(\m{S_R}\cdot \hat R)(\m{S_R'}\cdot \hat R') ]|\m R-\m R'|^{-3},\\
H_Z&=\mu_B \m H\cdot \sum \m {S_R}.\\
\end{aligned}	
\end{equation}
It is convenient to represent these expressions in $\m q$ space
\begin{equation}
\begin{aligned}
H_{EX}&=\frac{1}{2}\sum J_{\m q}\, \m{S_q\cdot S_{-q }},\\
H_{DM}&=\sum i\, D_\m q	\,\m{q\cdot [S_ q\times S_{-q}]},\\
H_{AE}&=\frac{1}{2}\sum_{\nu=x,y,z} F_\m q\, q^2_\nu\, S_\m q ^\nu\, S_{-\m q}^\nu,\\
H_D&=\frac{\omega_0}{2}\sum (\m S_\m q \cdot \hat q)(\m S_{-\m q}\cdot \hat q),\\
H_Z&=N^{1/2}\,\m H\cdot\m S_0.
\end{aligned}
\end{equation}
where $N$ is total number of the cells, $\m{S_q}=N^{-1/2}\sum \m S_\m R\exp (i\m{q\cdot R})$, $\hat q=\m q/q$, $ \omega_0=4\pi (g\mu_B)^2/v_0$ is the characteristic energy of the dipole interaction and  $v_0$ is the unit cell volume and  we omitted the isotropic part of the dipole interaction which has the same symmetry as the exchange one. The DMI and AEI are of the first and second order of the spin-orbit interaction respectively. So we have $J\gg D\gg F$. Value of the dipole energy $\omega_0$ will be given in the last Sec..

  \section{Classical energy}
Replacing $\m S_\m R$ by $S \ze$ we get the classical ground-state energy per unit cell in the following form
\begin{equation}
\begin{aligned}
E_{cl}&=-\frac{S^2}{2}(J_0\sin^2\alpha+J_\m k\cos^2\alpha)-S^2\,D_0(\m{k\cdot [\hat a\times \hat b]})\cos^2\alpha
+\frac{S^2 F}{4}[k_x^2(\hat a^2_x+\hat b^2_x)\\&+k_y^2(\hat a^2_y+\hat b_y^2)+k_z^2(\hat a_z^2+\hat b_z^2)]\cos^2\alpha+S h_\parallel \sin\alpha +\frac{S^2}{2} \omega_0  N_{cc}\sin^2\alpha,\\
\end {aligned}
\end{equation} 
here $h_\parallel=g\mu H_\parallel$, $H_\parallel$ is the field component along the $c$-axis and $N_{cc}$ is the corresponding component of the demagnetization tensor \cite{demag}. It is important to note that the classical energy depends on the field along $c$ axis only. Meanwhile experiment shows that the system is very sensitive to  weak perpendicular field \cite{leb}, \cite{koy} and \cite{ok}. We will explain below the nature of this quantum phenomenon.

  We are interested  by small $k$ only ($k a\ll1$) when  $J_\m k=J_0-\frac{A k^2}{S}$ and $D_\m k=D_0$. As we will see below $A$ is the spin-wave stiffness constant. At $H_\parallel=0$ this ground state energy was studied in [4].  From Eq.(6) for  components of the vector $\m k$ we get
\begin{equation}
A k_\nu+\frac{S F}{2}k_\nu (\hat a^2_\nu+\hat b^2_\nu)=S D_0 [\hat a\times \hat b]_\nu.
\end{equation}
from this expression we obtain
\begin{equation}
	A k^2+\frac{SF}{2}I(\m k)=S D_0(\m k\cdot[\hat a\times\hat b]),
\end{equation}
  where  $I=\sum k^2_\nu (a^2_\nu+b^2_\nu)$ is a cubic invariant. Inserting  $k_\nu$ from Eq.(7) into Eq.(6) we get
  \begin{equation}
	E_{cl}=\left\{-\frac{(S^3 D^2_0}{2A}+\frac{S^3F^2I^2}{8}+\frac {S^2 F I}{4}\right \}\cos^2\alpha+S h_\parallel \sin\alpha+\frac{S^2N_{cc}}{2}\omega_0\sin^2\alpha,
\end{equation}
where  the $F^2$ term may be omitted.
   There are two possibilities \cite{bak}. For negative $F$ the classical energy (6) is minimal  for $\m k$ along one of the cubic diagonals and $I=2 k^2/3$. If  $F>0$  the minimum is  for $\m k$ along the cubic edge and $I=0$. In both cases the spins rotate in the plane perpendicular to the vector $\m k$ and
 \begin{equation}
\m k=\frac{S D_0[\hat a\times \hat b]}{A+S F I/2},	
\end{equation}
 For $D_0>0$ and $D_0<0$ we have the right and left-handed helix respectively. 
The classical energy depends on the field projection onto  the vector $\m k$ ($c$-axis) only and  from Eq.(6) for  we obtain
\begin{equation}
\sin\alpha=
\begin{cases}
	-H_\parallel/H_c,& H_\parallel<H_c\\
	-1,& H_\parallel>H_c,
	\end{cases}
\end{equation}
where the critical field is given by
\begin{equation}
	g\mu_B H_c=h_c=A k^2+\frac{SF}{2}I+S\omega_0 N_{cc},
\end{equation}
here the last term is a result of the demagnetization and the intrinsic critical field is determined  by
 $H_c^{Int}=H_c-4\pi N_{cc} M $, where magnetization in the high-field "ferromagnetic" state  $M$  is determined as $4\pi\mu_B M=S\omega_0$  and  $g\mu H_c^{Int}=h _c^{Int}=A k^2+SFI/2\simeq A k^2$. For $H>H_c$ we have ferromagnetic spin configuration, but according to Eqs.\,(2) and (3) rotation persists in the perpendicular, spin-wave components of the spin density. 
 
\section{Spin-waves} 
We are interested in the long-wave spin-waves with $q\lesssim k$. At  $q\gg k$ one can neglect all interactions exept the exchange one and  the spin-wave spectrum must be the sane as in ferromagnets: $\epsilon_\m q=A q^2$. In the low-q region the hierarchy of the interactions becomes very important. The ferromagnetic exchange $J$ is the strongest one. The DMI and AEI are result of the weak  spin-orbit coupling  $\lambda$.  The former and the last are of order $\lambda J$ and $\lambda^2 J$  respectively. According to Eq.(10) $k\sim\lambda$.

 We are dealing with incommensurate spin structure where umklapp processes are important which mix excitations with momenta $\m q$ and  $\m{q\pm k}$ and different energies.It is useful to represent the unit vectors given by Eq.(2) as
\begin{equation}
	\begin{aligned}
	\ze&=\hat c \sin\alpha+(\m A e^{i\m {k\cdot R}}+\m A^*e^{-i\m{k\cdot R}})\cos\alpha,\\
	\et&=i\m A e^{i\m{k\cdot R}}-i\m A^*e^{-i{\m k\cdot R}},\\
	\ks&=\hat c\cos\alpha-(\m A e^{i\m{k\cdot R}}+\m A^* e^{-i\m{k\cdot R}})\sin\alpha	
	\end{aligned}
\end{equation}
  where $\m A=(\hat a-i\hat b)/2$, $\m{A\cdot A=A^*\cdot A^*=0}$, $\m{A\cdot A^*}=1/2$,  $\m A\times\hat c=-i\m A$, $\m A^*\times\hat c=i\m A^*$, and $\m A\times \m A^*=i\hat c/2$. Using this representation we obtain $\m{S_q}=S^c_\m q \,\hat c+S^A_\m q \m A+S_\m q^{A^*} \m A^*$ where 
\begin{equation}
	\begin{aligned}
	S^c_\m q&=S^\zeta_\m q  \sin\alpha+S^\xi _\m q  \cos\alpha,\\
	S^A_\m q&=S^\zeta_\m{q-k}\cos\alpha-S^\xi_\m{q-k}\sin\alpha+i S^\eta_\m{q-k},\\
	S^{A^*} _\m q&=S^\zeta_\m{q+k}\cos\alpha-S^\xi_\m{q+k}\sin\alpha-i S^\eta_\m{q+k}.
	\end{aligned}
\end{equation}
In the DMI  interference  terms appear proportional to   $S^c_\m q S^A_\m{-q-k}$ and $S^c_\m q S^{A^*}_\m{-q+k}$. They are responsible for umklapps with $\m{q\rightarrow q\pm k}$. At the same time the processes with $\m q\pm 2\m k$ appear in the $S^A S^A$ and $S^{A^*}S^{A^*}$ terms . Due to algebra of the vectors  $\hat c, \m A\,\mbox{and}\, \m A^*$ these double umklapps are in $H_{AE}$ and $H_D$ only. Besides as we  see below the perpendicular field initiate  processes with $\m{q\pm k}$ too.

 Here we evaluate the direct contribution to the spin-wave energy. The umklapps  will be considered in the next section.
 
 The exchange energy does not contain the umklapps and has the  form
\begin{equation}
\begin{aligned}
	H_{EX}&=-\frac{1}{2}\sum [(J_\m q \sin^2\alpha+J_{\m{q,k}}\cos^2\alpha)S^\zeta_\m q S^\zeta_{-\m q}+J_{\m{q,k}} S^\eta_\m q S^\eta_{-\m q}+(J_\m q\cos^2\alpha+J_{\m{q,k}}\sin^2\alpha) S^\xi_\m q S^\xi_{-\m q}\\&+i(S^\eta_\m q S^\zeta_{-\m q}-S^\zeta_\m q S^\eta_{-\m q}) N_{\m{q,k}}\cos\alpha+(J_\m q-J_{\m{q,k}})(S^\zeta_\m q S^\xi_{-\m q}+S^\xi_\m q S^\zeta_{-\m q})\sin\alpha\cos\alpha-i(S^\eta_\m q S^\xi_{-\m q }-S^\xi_\m q S^\eta_{-\m q}) N_{\m{q,k}}\sin\alpha ], 
\end{aligned}	
\end{equation}
where $J_{\m {q,k}}=(J_{\m{q+k}}+J_{\m{q-k}})/2\simeq J_0-A(q^2+ k^2)/S$ and $N_{\m {q,k}}=(J_{\m{q+k}}-J_{\m{q-k}})/2\simeq-2A\m{q k}/S$.\\
 Direct part of  $H_{DM}$ is given by
\begin{equation}
\begin{aligned}
H^{d}_{DM}&=-D_0(\m k[\hat a\times\hat b])\sum [S^\zeta_\m q S^\zeta_{-\m q}\cos^2\alpha+S^\eta_\m q S^\eta_{-\m q}+S^\xi_\m q S^\xi_{-{\m q}}\sin^2\alpha-(S^\zeta_\m q S^\xi_{-\m q}+S^\xi_\m q S^\zeta_{-\m q})\sin\alpha\cos\alpha]\\&+i D_0\sum(\m q [\hat a\times\hat b])[(S^\zeta_\m q S^\eta_{-\m q}-S^\eta_\m q S^\zeta_{-\m q})\cos\alpha +(S^\eta_\m q S^\xi_{-\m q}-S^\xi_\m q S^\eta_{-\m q})\sin\alpha],
\end{aligned}	
\end{equation}
where we neglected $\m q$-dependence of the DM interaction. For the  anisotropic exchange we have
\begin{equation}
	\begin{aligned}
H^{d}_{AE}&=\frac{F}{2}\sum_{\m q,\nu=x,y,z}\{(c_\nu q_\nu)^2(S^\zeta _\m q\sin\alpha+S^\xi_\m q \cos\alpha)(S^\zeta_\m{-q}\sin\alpha+S^\xi_\m{-q}\cos\alpha)+
\frac{(q^2_\nu+k^2_\nu)(\hat a^2_\nu+\hat b^2_\nu)}{2}\\&[(S^\zeta_\m q\cos\alpha-S^\xi_\m q\sin\alpha)(S^\zeta_\m{-q}\cos\alpha-S^\xi_\m{-q}\sin\alpha)+S^\eta_\m q S^\eta_\m{-q}]-2i q_\nu k_\nu( a^2_\nu+ b^2_\nu)(S^\zeta_\m q\cos\alpha-S^\xi_\m q\sin\alpha) S^\eta_\m{-q}\},
	\end{aligned} 
\end{equation}
In the case of the dipole interaction we obtain
\begin{equation}
	\begin{aligned}
	H_D^{(d)}&=\frac{\omega_0}{2}\sum \{(\hat q\cdot \hat c)^2(S^\zeta_\m q\sin\alpha+S^\xi_\m q\cos\alpha)(S^\zeta_\m{-q}\sin\alpha+S^\xi_\m{-q}\cos\alpha)\\&+\frac{q^2_\perp}{2|\m{q,k}|^2}[(S^\zeta_\m q\cos\alpha-S^\xi_\m q\sin\alpha)(S^\zeta_\m{-q}\cos\alpha-S^\xi_\m{-q}\sin\alpha)+S^\eta_\m q S^\eta_\m{-q}] \},
	\end{aligned} 
\end{equation}
where $q^2_\perp=q^2_a+q^2_b$  and  $|\m{q,k}|^{-2}=(|\m{q+k}|^{-2}+|\m{q-k}|^{-2})/2$. In  this expression we have taken into account that the vector $\m k$ is perpendicular to  the $a b$ plane.\\
Replacing in Eqs.(15)-(18) $S^\zeta_\m q$ by $N^{1/2} S\delta_{\m q,0}$ we obtain a sum of terms proportional to $S^\xi_0$ which cancels corresponding contribution from the Zeeman energy. In the sum $H_{EX}+H^d_{DM}+H^d_{AE}$ the terms proportional to  $\sin\alpha\, \mbox{and}\, \cos\alpha$ cancel too due to Eq.(8), if one neglects small terms of order of $F \m k\cdot\m q$. 

We consider now linear spin-wave theory. From Eqs.\,(3),\,and (15)-(18) follows
\begin{equation}
	H^{(d)}_{SW}=\sum[E_\m q a^+_\m q a_\m q +\frac{B_\m q}{2}(a_\m q a_{-\m q}+a^+_\m q a^+_{-\m q})],
\end{equation}
where 
\begin{equation}
\begin{aligned}
E_\m q&=A q^2+(\frac{A k^2}{2}+\frac{S F I}{2})\cos^2\alpha+\frac{S\omega_0}{2}[\hat q^2_c\cos^2\alpha\\&
+\frac{q^2_\perp}{|\m{q,k}|^2}(1+\sin^2\alpha)]-S(h_c\sin\alpha+h_\parallel)\sin\alpha ,
\end{aligned}
\end{equation}
where $h_\parallel=g\mu_B H_\parallel$ and $h_c$ is determined by Eq. (12). Using Eq.(11) one can show that expression in the last  brackets is zero if $H_\parallel<H_c$. For $B_\m q$ we have   
\begin{equation}
\begin{aligned}
B_\m q=(\frac{A k^2}{2}+\frac{S F I}{2})\cos^2\alpha+\frac{S \omega_0}{2}(\hat q^2_c-\frac{q^2_\perp}{|\m{q,k}|^2})\cos^2\alpha,
\end{aligned} 	
\end{equation}
 The spin-wave energy is determined as $\epsilon_\m q=(E^2_\m q-B^2_\m q)^{1/2}$. For $H_\parallel>H_c$ we have 
\begin{equation}
\epsilon_\m q=A q^2+h_\parallel-h_c+\frac{S\omega_0 q^2_\perp}{|\m{q,k}|^2},
\end{equation}
At $H_\parallel<H_c$ the energy has the form	
\begin{equation}
\epsilon_\m q=\left\{\left(A q^2+\frac{S\omega_0 q^2_\perp}{|\m{q,k}|^2}\right)\left[A q^2+(A k^2+S\omega_0\hat q^2_c)\cos^2\alpha+\frac{S\omega_0 q^2_\perp }{|\m{q,k}|^2 }\sin^2\alpha \right]\right\}^{1/2}
\end{equation}
 and we have the gapless excitation with linear dispersion at $q\ll k$. In these equations the factor $q^2_\perp/|\m{q,k}|^2$ is singular if $\m{q\rightarrow\pm k}$ and should be replaced by $\hat q^2_\perp/2$ or $N_\perp/2=(N_{a a}+N_{b b})/2$ if  $\m q_\perp =0$. This replacement becomes evident if we put $\m{q=k+q_\perp}$.

  \section{Umklapp interaction}
  There are following umklapp contributions to the Hamiltonian: i) Interaction of the perpendicular magnetic field with $S^\zeta$ spin component. ii)Interference of the longitudinal ($c$) and transverse ($\m A$ and $\m A^*$) spin components in the DM and dipole interactions. All these contributions mix excitations with $\m q$ and $\m{q\pm k}$. The dipole interaction contains also the $(\m {A\, A})$ and $(\m{A^*\, A^*})$ terms which mix waves with $\m q$ and $\m{q\pm 2k}$. We will show below that this contribution may be neglected. The umklapps in the anisotropic exchange is neglected too.
  
  The umklapp contribution to the Hamiltonian consists from terms odd and even in operators $a$ and $a^+$. It is convenient  to write out general expression  for it and then extract the bilinear part which contribute to the spin-wave spectrum and linear one considered in the next section. So we have
\begin{equation}
	\begin{aligned}
	H_Z^{(U)}&=-\m{h\cdot A}(S^\xi_\m{-k}\sin\alpha-i S^\eta_\m{-k}+\sum a^+_\m{q+k}a_\m q\cos\alpha)\\&-\m{h\cdot A^*}(S^\xi_\m k \sin\alpha+i S^\eta_\m k+\sum a^+_\m{q-k}a_\m q\cos\alpha),\\
H_U&=\sum \{[-2D_0(\m{q\cdot A})+\omega_0(\hat c\cdot \hat q)(\hat q\cdot \m A)] S^c_\m q S^A_\m {-q}\\&+[2D_0(\m{q\cdot A^*})+\omega_0(\hat c\cdot\hat q)(\hat q\cdot \m A^*)]S^c_\m q S^{A^*}_\m { -q}\},\\
	\end{aligned}
\end{equation}
 where $H_U=H^{(U)}_{DM}+H^{(U)}_D$, and $S^c$, $S^A$ and $S^{A^*}$ are given by Eq. (14). In this expression we included also terms linear in $S^{\xi,\eta}_\m{-k}$. They are not contribute to the spin-wave energy and will be considered in next section.
 
 The term $S\delta_{\m q,0}$ in the expression for $S^c_\m q$ leads to demagnetization of the perpendicular field if $H_\parallel\neq 0$ which may appear due to low symmetry of the sample. In this case  $-\m h\cdot\m A$ is replaced by
 \begin{equation}
  P=-(\m h\cdot \m A+S\omega_0 N_{c A}\sin\alpha),
  \end{equation}
   where $N_{c A}=(N_{c a}-i N_{c b})/2$ is corresponding off-diagonal component of the tensor $N_{\alpha\beta}$ \cite{demag}. It is an unusual feature which has to be clarified. The tensor $N_{\alpha\beta}$ is diagonal if the coordinates are along the principal axes or the ellipsoidal sample. If $\m k$ is not along one of these axes   the off diagonal components of $N_{\alpha\beta}$ appear.   The $\delta_{\m q,\pm\m k}$ terms in $S^{A, A^*}_\m q$ do not contribute due to the condition $\m q\cdot\m A=0$. As a result for the bilinear spin-wave contribution we  obtain
\begin{equation}
	\begin{aligned}
	H^{(U)}_{SW}&=\sum\{[P\cos\alpha+R_\m{-q-k}(1-\sin\alpha)\\&-R_\m q(1+\sin\alpha)] a^+_\m{q+k}a_\m q+[P^*\cos\alpha\\&+R^*_\m{-q}(1-\sin\alpha)-R^*_\m{q-k}(1+\sin\alpha)] a^+_\m{q-k}a_\m q\\&+
	R_\m q[a_\m q a_\m{-q-k}(1-\sin\alpha)-a^+_\m{-q}a^+_\m{q+k}(1+\sin\alpha)]\\&+R^*_\m{-q}[a^+_\m{-q}a^+_\m{q-k}(1-\sin\alpha)-a_\m q a_\m{-q+k}(1+\sin\alpha)]\}
	\end{aligned}
\end{equation}
 where 
\begin{equation}
	R_\m q=\frac{S\cos\alpha}{2}[-2D_0(\m q\cdot\m A)+\omega_0(\hat c\cdot\hat q)(\hat q\cdot\m A)].
	\end{equation}
Here again  $R_\m q$ contains off diagonal component of $N_{\alpha\beta}$.	From these equations we see that the umklapp interaction contributes to the spin-wave energy at $H_\parallel<H_c$ only and do not affect excitations with $\m q$ along $\m k$. At the same time from Eqs. (15)-(18) for the direct interaction follows that at $H_\parallel >H_c$ the zero-point vibrations disappear.
	
For consideration of the umklapp interaction  we will use   equations of motion
\begin{equation}
 \omega G_{A,B}(\omega)+G_{[H,A],B}(\omega)=<[A,B]>,	
\end{equation}
where $H=H^{(d)}_{SW}+H^{(U)}_{SW}$ and the Green functions are determined by
\begin{equation}
	G_{A,B}(\omega)=-i\int_{0}^{\infty}  e^{i \omega t}<[A(t),B(0)]>=-<A,B>_\omega
\end{equation}	
where the expression in the right hand side  coincides with the conventional determination of the generalized susceptibility. 

We will use below the following notations
\begin{equation} 
\begin{aligned}
G_{a_\m q,a^+_\m q}&=G_\m q, &G_{a_{\m{q\pm k}},a^+_\m q}&=G_{\pm 1},\\
F_{a^+_\m{-q},a^+_\m q}&=F_\m q, & F_{a^+_{\m{-q\mp k}},a^+_\m q}&=F_{\pm 1}.
\end{aligned}
\end{equation}
Neglecting the umklapp interaction  we have 
\begin{equation}
	G_\m q(\omega)=\frac{E_\m q+\omega}{\omega^2-\epsilon^2_\m q};\quad F_\m q(\omega)=-\frac{B_\m q}{\omega^2-\epsilon^2_\m q},
\end{equation}
where  $E_\m q$, $B_\m q$ and  $\epsilon_\m q$ are given by Eqs. (20)--
 (23). If $H_\parallel>H_c$ we have $F=0$\\
Using Eq.(28) we obtain infinite set of equations which contain along with conventional Green functions $G$ and $F$ the functions $G_{\pm n\m k}$ and $F_{\pm n\m k}$ where $n=1,2\ldots$. Below we will restrict mainly by $n=1$ approximation. Its validity  will be discussed later. As a result we obtain six linear equations for functions (30) (see Appendix A). In general form  their solution is very complex. So we will consider two limiting cases: $q\rightarrow 0$ and $q\gtrsim k$.

 \underline{\textit{The  $\m q\rightarrow 0$ case}}. Corresponding equations are analyzed in Appendix B. Neglecting the dipole interaction we obtain the final result for weak perpendicular field in the form 
\begin{equation}
\begin{aligned}
	G_\m q (\omega)&=\frac{(A k^2/2)\cos^2\alpha+\omega}{\omega^2-\epsilon^2_\m q+ (h_\perp^2/2)\cos^4\alpha},\\
	F_\m q(\omega)&=-\frac{(A k^2/2)\cos^2\alpha}{\omega^2-\epsilon_\m q^2+ (h^2_\perp /2)\cos^4\alpha},
	\end{aligned}
\end{equation}
where $h_\perp^2=(g\mu_B)^2(H_a^2+H_b^2)$. We neglected here the small terms  proportional to $h^2_\perp$ in numerators as well as the contribution of the slightly splitted modes with the energies  $\epsilon\simeq \epsilon_\m k$ (See Appendix B). The functions  $G_\pm$ and $F_\pm$ are of order of  $h_\perp $ and are evaluated in Appendix B.
In these expressions  the square of the spin-wave energy in the transverse field at $\m q=0$ is given by Eq.(B5)
\begin{equation}
	\epsilon^2_\m q(H_\perp)=\Delta^2-(h^2_\perp/2) \cos^4\alpha.
\end{equation}
 For the gapless spin-waves  $\epsilon^2_0 (H_\perp)$ is negative for any $H_\perp$ and magnetic subsystem is unstable. It means that the helix axis $\m k$ must turn and stands along the field.   For finite  gap $\Delta$, the spin-wave spectrum remains stable up to $h_\perp \sim\Delta$ and then  there is the first order transition to the parallel state. In the intermediate case, when $H_\parallel\neq 0$ there is rather complex behavior governed by the equation of state  derived  in the next Sec. Similar situation takes place in conventional antiferromagnets: If the field is along the sublattice magnetization  the first order spin-flop transition occurs at $H=\Delta$ where $\Delta$ is the spin-wave gap. For the inclined field there is the rotation of the sublattices along with the first order transition \cite{mal}. We consider  origin of the gap  below.
 
 The rotation of the helix axis was observed in $F e G e$ \cite{leb} and $M n S i$ \cite{koy} and \cite{ok}. In both cases $H_\perp$ is much less than $H_c$. For  $M n S i$ the transition to parallel state is at   
  $H_{\perp c} \sim 0.1T$ and $H_c\simeq 0.6T$. It is the reason of the approximation with $n=1$. Any additional  $n$ produces small correction of order of $H^2_\perp/H^2_c\ll 1$.
  
  \underline{\textit{The $q\gtrsim k$ case}}.
   As above we consider the first-order umklapps  connecting $\m q$ and $\m{q\pm k}$. We consider $\m q\perp \m k$ case only  where the umklapp interaction  the dipole interaction is absent (see Eq.(27). We put also $\m H=0$. Then we discussed briefly the role of small $\m q_\parallel$. We demonstrate also that $n=2$ approximation  changes  does not change qualitatively  the $n=1$ results.
   
   If $\m q\perp \m k $  we have two interacting modes with  $\m q$  and  $\m{q\pm k}$  which energies according to Eq. (23) are given by
\begin{equation}
	\begin{aligned}
	\epsilon_\m q&=A q_\perp(q_\perp^2+k^2)^{1/2}\\
	\epsilon_{\m q\pm k}&=A[(q_\perp^2+k^2)(q_\perp^2+2k^2)]^{1/2}.
	\end{aligned}
\end{equation}
  The excitation $\epsilon_\m q$ is the gapless and linear in $\m q$ at $q\ll k$. The gap of the double degenerated mode is equal to $A k^2\sqrt{2}$. The umklapp interaction lifts this degeneracy  (see Appendix C) and we have three different modes. One of them remains non-renormalized and coincides with the gapped mode in Eq.(34) and two other are given by
\begin{equation}
	\epsilon_\pm=A[k^2+4q_\perp^2+q_\perp^4/k^2 \pm(k^2+8q_\perp^2+17q_\perp^4/k^2+8q_\perp^6/k^4)]^{1/2},
	\end{equation}
	where  the plus mode has the same gap as in Eq.(34) but other dependence on $q_\perp$. 
	
	The non-renormalized mode is irrelevant as corresponding expressions for the Green functions given by Eqs.(C5) do not have poles at this energy. It is evident that for other directions of  $\m q$  the Green functions must have all three poles. The gapless mode is strongly renormalized and at small $q_\perp$ we have quadratic dispersion with $\epsilon_-=A q^2_\perp/2$. Asymptotically at  $q\gg k$ all three modes tends to $A q^2$ and the umklapps become unimportant.  The umklapp renormalization is seen clearly in Fig. 1  where the full and dashed lines correspond to minus and plus  modes in Eq(35) and the dash-dotted line to gapless mode with $ \ q\parallel \m k$ respectively.   
  
  If $\m q\parallel\m k$ the umklapp interaction  disappears (see Eq. (27)). General expression for arbitrary directed $\m q$ is very complex. But it becomes very simple  for small $q_{\perp,\parallel}\ll k$. In this case from Eq. (C3) we get
\begin{equation}
	\epsilon_\m q=A[q^4_\perp/2+k^2 q_\parallel^2]^{1/2}.
\end{equation}
Cancellation of the $q^2_\perp$ term is the most striking feature of this expression. So we have  to discus validity of our $n= 1$ approximation. Corresponding equations of motion are given by Eq. (A1) 
where at $H=0$ and $\m q=\m q_\perp$ and the  amplitude of the umklapps $R_\m q=-SD_0 (\m q\cdot \m A)$. So at small $q_\perp\ll k$  and $\omega\ll A k^2$ we have an expansion in powers of $q^2_\perp$ as $G_\pm$ and $F_\pm $ are proportional to $1/\epsilon_\m k$. So the $n=1$ approximation  gives  correct result for $q^2_\perp$ term at small $\omega$ and  above cancellation of $q_\perp^2$  survives in  higher approximations. It shown in Appendix C for $n=2$. However the $q_\perp^4$ remains with slightly changed coefficient and instead Eq.(36) we obtain
\begin{equation}
	\epsilon_\m q=A[3q^4_\perp/8 +k^2q^2_\parallel]^{1/2}
	\end{equation}
	It is follow from the above argumentation that further approximations can not change coefficients in this expression. For $q_\perp\gg k$ we have  $1/q_\perp$ expansion and the umklapp renormalization disappears. Unfortunately at $q_\perp\sim k$ and $\omega\sim A k^2$ the results of $n=1$ approximation are qualitatively correct only. It is demonstrated  in the end of Appendix C.
	
	\section{Perpendicular susceptibility and helix deformation.}
We have demonstrated that the terms linear in operators $a$ and $a^+$ cancel in the diagonal part of the interaction given by Eqs. (15)-(18) due to equilibrium conditions for the classical energy. However there are not similar conditions  for the terms with $a_{\pm\m k}$ and $a^+_{\mp\m k}$ in  the interaction (24) with the perpendicular field and we have
\begin{equation}
	H^U_Z=P\left(\frac{S}{2}\right)^{\frac{1}{2}}[-a_{-\m k}(1-\sin\alpha)+a^+_\m k (1+\sin\alpha)]+h.c.
\end{equation}
This expression  may be considered  as perturbation which gives rise addition contribution in the energy as well as  some non-zero parts of the average values of the transverse spin components $<S^{\xi,\eta}_\m R>$ proportional to $\m h_\perp$. But we will use more general method, which leads in the first approximation to the same results. We consider  operators $a_{\pm \m k}$ and $a^+_{\pm \m k}$ as c-numbers. Corresponding c-number terms appear in Eqs.(19) and (26) and we obtain additional contribution to the ground-state energy 
\begin{equation}
\begin{aligned}
	E_1&=P\sqrt{S/2}[a^+_\m k(1+s)-a_\m{-k}(1-s)]+[P c-R_0(1+s)]a^+_\m k a_0\\&
	[P c+R_0(1-s)]a^+_0 a_\m{-k}+ R_0[a_0 a_\m{-k}(1-s)-a^+_0 a^+_\m k (1+s)]+h.c.\\&
	+E_\m k(|a_\m k|^2+|a_\m{-k}|^2)+B_\m k(a_\m k  a_\m{-k}+a^+_\m k a^+_\m{-k}) +E_0|a_0|^2+B_0 (a^2_0+a^{+2})/2, 
	\end{aligned}
\end{equation}
where $s=\sin\alpha$, $c=\cos\alpha$, $R_0=(S/2)\omega_0 N_{c A}\cos\alpha$ and  in the classical limit  $a^+=a^*$.

The energy $E_1$ is minimal at the following  obvious equilibrium conditions $\partial E_1/\partial a^{(+)}_{0 ,\pm\m k}=0$ which lead again in $n=1$ approximation to six linear equations. Corresponding matrix coincides with $M$ given by Eq.(B2) at $\omega=0$ and  we get equation 
\begin{equation}
	M(0)Z=V,
\end{equation}
where column $Z=(a_0,a^+_0,a_\m{-k},a^+_\m k,a_\m k,a^+_\m{-k})$ and
 \begin{equation}
 V=(0,0,-P^*\sqrt{S/2}(1-s),-P^*\sqrt{S/2}(1+s),P\sqrt{S/2}(1+s),P\sqrt{S/2}(1-s)).	
\end{equation}
In the case when dipole contribution $R_0=0$ solution of these equations is presented in Appendix B. Using above-mentioned equilibrium conditions one can show
\begin{equation}
	E_1=\frac{P}{2}\sqrt{\frac{S}{2}}[(a_\m k+a^+_\m k)(1+s)-(a_\m{-k}+a^+_\m{-k})(1-s)]+c.c.,
\end{equation}
and  for the field depended part of the ground-state energy we have
\begin{equation}
	E=-\frac{Sh^2_\parallel}{A k^2}-\frac{S h^2_\perp \Delta^2}{2A k^2(1+\cos^2\alpha)[\Delta^2-(h^2_\perp/2)\cos^4\alpha]},
\end{equation}
where the first term is the Zeeman energy for $H<H_C$.

The helix axis has to rotate at $g\mu_B H_\perp\sim\Delta$. 
According to experimental data \cite{leb}, \cite {ok} the rotation was observed  at  $H_\perp \ll H_c$. So we can put $\cos^2\alpha=1$ and obtain
\begin{equation}
	E=\frac{Sh^2}{A k^2}\left[-\cos^2\Psi-\frac{\\sin^2\Psi}{4(1-h^2_\perp \sin^2\Psi/(2\Delta^2)}\right],
\end{equation}
where $\Psi$ is the angle between the helix axis $\m k$ and the field. For small $h_\perp$  this energy is minimal at $\Psi=0$ i.e the field along  $\m k$. The real direction of the helix axis is determined by competition between the magnetic energy and anisotropic exchange given by Eq. (9). Eq.(44) is not valid at $h\sim \Delta $ as it was derived in the linear approximation when the amplitudes $a$ and $a^+$  given by  (B7) should be small.

The perpendicular field deforms the helix structure producing the higher harmonics. In our $n=1$ approximation there is the second harmonic only. The static contribution to the transverse spin components  has now the form
\begin{equation}
	\m{S_R}=i\m A e^{i \m{k\cdot R}}(S^\eta_\m k e^{i {\m k\cdot R}}+S^\eta_\m{-k}e^{-i {\m k\cdot R}}]+c.c.,
	\end{equation}
where $i S^\eta_\m{\pm k}=\sqrt{S/2}(a_\m{\pm k}-a^+ _\m{\mp k})$ and we put $\alpha=0$. From these expressions using Eqs. (B7) we obtain the transverse uniform magnetization \cite{def}
\begin{equation}
	\m S^\perp_U=-\frac{S \m h_\perp\Delta^2}{2A k^2(\Delta^2-h^2_\perp/2)}
\end{equation}
The second harmonic of the helix structure induced by the perpendicular field is given by
	\begin{equation}
\m S^\perp_\m R=-\frac{2\Delta^2 }{2A k^2(\Delta^2-h^2_\perp/2)}[\m{A(A\cdot h)}e^{2i\m{k\cdot R}}+\m{A^*(A^*\cdot h)}e^{-2i\m{k\cdot R}}]	
\end{equation}
As we have mentioned above corresponding second-order Bragg reflections were observed \cite{leb} and \cite{ok}.

\section{Cubic anisotropy and the gap problem}
We postulated above existence of the spin-wave gap. Now we demonstrate that there are at least two contributions to the gap: cubic anisotropy and interaction between spin-waves considered in the Hartree-Fock (HF) approximation. The former may have arbitrary sign and the last is positive. So  different contributions to the gap may compete. Changing the sign and strength of the cubic anisotropy by pressure one can get the quantum phase transition from the ordered to the spin-liquid state. This  phenomenon was observed in $M n S i$ \cite{pf1}, \cite {pf2} and \cite{f}.\\
Complete expression for the gap at arbitrary $H_\parallel$ is very complex. So we present here results for $H=0$ and $H>H_c$. 
We begin with the cubic anisotropy.  As is well known the single ion cubic anisotropy has the form
\begin{equation}
	V=K\sum_\m R(S^4_{x,\m R}+S^4_{y,\m R}+S^4_{z,\m R}).
\end{equation}
Using Eqs. (3), (13) and (14) for zero magnetic field  ($\alpha=0$) we obtain
\begin{equation}
	V=6S^4K\sum_{\nu=x,y,z}|A_\nu|^4+6S^3K\sum_{\m k,\nu=x,y,z}\{2|A_\nu|^2\hat c^2_\nu a^+_\m k
a_\m k+[|A_\nu|^2\hat c^2_\nu-2|A_\nu|^4](a_\m k a_\m{-k}+a^+_\m k a^+_\m{-k})\}.
\end{equation}
 Neglecting in Eq. (9) the $F^2$ term for the classical energy  we have
\begin{equation}
	E_{cl}=\frac{S^2F}{4}\sum_{\nu=x,y,z}k^2_\nu(\hat a^2_\nu+\hat b^2_\nu)+\frac{3S^4 K}
{8}\sum_{\nu=x,y,z}(a^2_\nu+b^2_\nu)^2.
\end{equation}
The spin-wave terms in Eq.(49) give additional contribution to $E_\m q$ and $B_\m q$ in Eq.(19) which are given by
\begin{equation}
	\delta E_\m q=12S^3 K\sum \hat c^2_\nu |A_\nu|^2;\quad\delta B_\m q=12S^3K\sum(\hat c^2_\nu |A_\nu|^2-2|A_\nu|^4), 
\end{equation}
 and for the  anisotropic contribution to the spin-wave gap we obtain
\begin{equation}
	\Delta^2_c=\frac{3}{2} S^3 K h_c\sum_{\nu=x,y,z}(\hat a^2_\nu+\hat b^2_\nu)^2,
	\end{equation}
	where $h_c=E_0+B_0$, $h_c$, $E_0$ and $B_0$ are given by Eqs. (12), (20) and (21). This equation for the gap holds   at $\m q=0$ when $\hat q^2_c\rightarrow N_{cc}$. For small $\m q$ we must replace $h_c$  by $h^{Int}_c+S\omega_0 \hat q^2_c$. So at $\m q=0$ and $\m q\neq 0$ the gap depends on the sample form and angle between $\m q$ and the helix axis respectively. We see that $\Delta^2_c$ is positive for $K>0$ only. In  ferromagnets this sign of $K$ corresponds to the easy directions along the cubic diagonals. 
	
  Let us consider now the classical energy and the gap for  orientations of the helix vector $\m k$ along $[1,1,1]$, $[1,0,0]$ and [1,1,0] directions labeled as 1, 2, and 3 respectively. We can choose $\hat c_1=(1,1,1)/3^{1/2}$, $\hat a_1=(1,-1,0)/2^{1/2}$ and $\hat b_1=(1,1,-2)/6^{1/2}$; $\hat c_2=(1,0,0)$, $\hat a_2=(0,1,0)$ and $\hat b_2=(0,0,1)$; $\hat c_3=(1,1,0)/2^{1/2}$, $\hat a_3=(1,-1,0)/2^{1/2}$ and $\hat b_3=(0,0,1)$. For the classical energy we have 
\begin{equation}
\begin{aligned}
E_{cl 1}&=\frac{S^2F k^2}{6}+\frac{S^4 K}{2};\quad E_{cl 2}=\frac{3S^4 K}{4};\quad E_{cl 3}=\frac{S^2F k^2}{8}+\frac{9 S^4 K} {16};
\end{aligned}
\end{equation}
From these expressions we see  that structures $(1,1,1)$ is realized if $S^2F k^2<3 S^4K/2$. Otherwise we have $(1,0,0)$ structure.  These expressions  are derived at $T=0$. However interplay between two anisotropy constants $F$ and $K$ must determine the helix direction at  all $T$ and explain the transition from $(1,0,0)$ to $(1,1,1)$ structure in $F e G e$ with decreasing $T$ \cite{leb}.

 Corresponding expressions  for the gap are given by
\begin{equation}
\begin{aligned}
\Delta^2_{c1}=4S^3K h_c;\quad \Delta^2_{c2}=6S^3K h_c;\quad \Delta^2_{c3}=(9/2)K S^3 h_c.  
\end{aligned}	
\end{equation}

We consider now  contribution to $\Delta^2$ arising from the spin-wave interaction considered in the HF approximation. It is well-known that in the Heisenberg ferromagnets there is not a gap due to the total spin conservation law. The DM interaction  brakes this law. From Eqs. (15) and (16) we have
\begin{equation}
	H_{EX}+H^d_{DM}=-\frac{1}{2}\sum\{[J_{\m{q ,k}}-J_0+2D_\m q (\m q\cdot[\hat a\times\hat b])](S^\zeta_\m q  S^\zeta_\m{-q}+S^\eta_\m q S^\eta_\m{-q})+(J_\m q-J_0)S^\xi_\m q S^\xi_\m{-q}\}
\end{equation}
where we have taken into account that $\sum\m{S_q S_{-q}}=NS(S+1)$ and the odd terms with $S^\zeta S^\xi$ surviving at  arbitrary $\m q$ are zero in the HF approximation. We replace also $D_0$ by ${D_\m q}$  as  the HF  contribution is saturated by large $\m q$.  Using now Eq.(3) for the  spin-wave interaction energy we obtain 
\begin{equation}
	V_I=-\frac{1}{2}\sum\{ V_\m 1 [a^+_\m{1+2} a_\m 2 a^+_\m{3-1}a_\m 3+\frac{1}{2}(a_\m 1-a^+_\m{-1})a^+_\m{1+2+3}a_\m 2 a_\m 3]-\frac{1}{2}(J_\m 1-J_0)(a_\m 1+a^+_\m{-1})a^+_\m{1+2+3}a_\m 2 a_\m 3\}, 
\end{equation}
 where  $\m{1,2,3=\m q_{1,2,3}}$ and $V_\m 1=J_\m{1,k}-J_0+2D_\m 1(\m{k\cdot[\hat a\times \hat b]})$. \\
 The HF contribution to  $\Delta^2$ is calculated in Appendix D by the method used in \cite{pet} for evaluation HF corrections to the spin-wave energy in 2D $C u O_2$ planes where the gap appearing due to pseudodipolar interaction between  neighboring $Cu^{2+}$ ions. From Eqs. (52) and (D7) we obtain the final result
\begin{equation}
	\Delta^2=\frac{3S^3 K h_c}{2}\sum(\hat a^2_\nu+\hat b^2_\nu)^2+\frac{A k^2 h_c }{4S}\sum\frac{D_\m q}{D_0},
\end{equation}\\ 
where the last sum has to be less than unity.

In  strong field  $H>H_c$ we have $\cos\alpha=0$ and zero-point fluctuation disappear ($B_\m q=0$ in Eq. (21)). As a result one can show that there is not the HF contribution to the gap and the classical energy  and the gap are  given by
\begin{equation}
\begin{aligned}
	E_{cl}&=\frac{S\omega_0 N_{cc}}{2}-S h_\parallel+S^4 K\sum \hat c^4_\nu\\&
	 \Delta=h_\parallel-h_c+4S^3 K\sum[-\hat c^4_\nu+\frac{3}{2}\hat c^2_\nu (\hat a^2_\nu+\hat b^2_\nu)].
	 \end{aligned}
\end{equation}
For the cases listed above cubic contributions to the classical energy and the  gap are given by
 \begin{equation}
	\begin{aligned}
	E_{cl1}&=\frac{K S^4}{3};\quad E_{cl2}=K S^4;\quad E_{cl3}\frac{K S^4}{2};\\&
	\Delta_1=\frac{8K S^3}{3};\quad \Delta_2=-4K S^3;\quad \Delta_3=
	 K S^3.\\ 
	\end{aligned}
\end{equation}

\section{EPR and Neutron scattering}   
Both phenomena the EPR and the inelastic magnetic scattering are described by the spin susceptibility. We outline here some of its  features.  For simplicity we use the $n=1$ approximation. 

From Eq.(13) and definition (29) we obtain following general expression for  the susceptibility
\begin{equation}
	\begin{aligned}
	\chi_{\alpha\beta}(\m q,\omega)&=<S^c_\m q,S^c_\m{-q}>_\omega \hat c_\alpha \hat c_\beta+<S^A_\m q,S^{A^*}_\m{-q}>_\omega A_\alpha A^*_\beta+<S^{A^*}_\m q,S^A_\m{-q}>_\omega A^*_\alpha A_\beta\\&+<S^c_\m q,S^A_\m{- q}>_\omega \hat c_\alpha A_\beta+<S^A_\m q,S^c_\m{-q}>_\omega A_\alpha\hat c_\beta+<S^c_\m q,S^{A^*}_\m{-q}>_\omega \hat c_\alpha A^*_\beta+<S^{A^*}_\m q,S^c_\m{ q}>_\omega A^*_\alpha \hat c_\beta\\&+
	<S^A_\m q,S^A_\m{-q}>_\omega A_\alpha A_\beta+<S^{A^*}
_\m q,S^{A^*}_\m{-q}>_\omega A^*_\alpha A^*_\beta,
	\end{aligned}
\end{equation} 
where $-<A,B>_\omega$ is the generalized susceptibility $chi_{AB}(\omega)$. In the first three terms both spin components have opposite momenta and describe the direct processes . All other are the umklapp terms. We will use below the linear spin-wave theory  and consider two limiting cases $H=0$ and  $H>H_c$.  The main attention will be paid to direct part of the susceptibility. The umklapp part will be  commented briefly in the end of this Sec.

\underline{\textit{Zero-field case}}.

There are  now two contributions to the susceptibility along the $\m k$
and in $ab$ plane. For the first using Eqs. (14) and (31) we have
\begin{equation}
\chi_{cc}(\m q,\omega)=-(S/2)[G_\m q(\omega)+G_\m q(-\omega)+F_\m q(\omega)+F^+_\m q(\omega)],	
\end{equation}
where due to non-Hermiticy  of the interaction (56) we must to distinguish between $F$ and $F^+$. In the EPR case when $\m q=0$ we obtain
\begin{equation}
	\chi_{cc}(\omega)=-\frac{S \Delta^2}{h_c(\omega^2-\Delta^2)}.
\end{equation}
 This susceptibility describes respond to the external ac field. Connection between external and intrinsic susceptibilities is given by well known equation (see for example \cite{mal3})
\begin{equation}
	\chi_{\alpha\beta}=\chi^{Int}_{\alpha\beta}-4\pi \omega_0 \chi^{Int}_{\alpha\mu}N_{\mu\nu}\chi_{\mu\beta},
\end{equation}
	and for $\chi^{int}_{cc}$ we obtain the same equation as for $\chi_{cc}$ with replacement $h_c$
and $\Delta^2$ by corresponding intrinsic quantities. It should mention that there has to be very small contribution of the $\epsilon_1$ mode (more precisely $\epsilon_+$ mode). It  appears if in Eqs. (B9-10) we retain terms proportional to $(E_0-B_0)|R|^2=\Delta^2|R|^2/(E_0+B_0)$. Moreover in $n=2$ approximation  the $\epsilon_2$  mode has to appear too. But we neglect both these contributions.

For $q\gtrsim k$ the susceptibility is strongly anisotropic in $q$ space. For  $\m{q\parallel k}$ the umklapp interaction is zero and from  we obtain 
\begin{equation}
	\chi_{cc}(q_\parallel,\omega)=-\frac{S A q^2_\parallel}{\omega^2-\epsilon^2_{q_\parallel}},
\end{equation}
where  $\epsilon^2_{q_\parallel}=A q^2_\parallel [A(q^2_\parallel+k^2)+S\omega_0 \hat q^2_\parallel]$.\\
 With increasing of $\m q_\perp$ tree different modes appear. However at $q_\parallel=0$  there are two modes (see Appendix C)  given by Eqs. (35) and using Eqs. (C5) we get
\begin{equation}
	\chi_{cc}(\m q_\perp,\omega)=-\frac{S A q^2_\perp}{\epsilon^2_+-\epsilon^2_-}\left[\frac{\epsilon^2_+-\epsilon^2_1+A(2q^2_\perp+k^2)}{\omega^2-\epsilon^2_+}+\frac{\epsilon^2_1-\epsilon^2_--A(2q^2_\perp+k^2)}{\omega^2-\epsilon^2_-}\right],
\end{equation}
 where  $\epsilon_1$ is given by the second line in Eq. (34). In  the $n=2$ approximation additional  modes  appear and so on. However amplitudes of these  modes  decrease  with $n$.
 
  Consider now the second two terms in Eq. (60). By the same way as above we get
\begin{equation}
	\chi^{A A^*}_{ \alpha\beta}(\m q,\omega)=(\delta_{\alpha\beta}-\hat c_\alpha\hat c_\beta)\chi_{\perp \m q}(\omega)-\frac{i}{2}\epsilon_{\alpha\beta\gamma}\hat c_\gamma C_\m q(\omega),
	\end{equation}
	where we put $\hat a_\alpha\hat a_\beta+\hat b_\alpha\hat b_\beta=\delta_{\alpha\beta}-\hat c_\alpha\hat c_\beta$ and $\hat a_\alpha\hat b_\beta-\hat a_\beta\hat b_\alpha=\epsilon_{\alpha\beta\gamma}\hat c_\gamma$.  Perpendicular and chiral susceptibilities are determined as
	\begin{equation}
	\chi_{\perp \m q}=\chi_\m{q+k}(\omega)+\chi_\m{q-k}(\omega);\quad C_\m q (\omega)=\chi_\m{q+k}(\omega)-\chi_\m{q-k}(\omega),
\end{equation}
where
\begin{equation}
\chi_\m Q(\omega)=\frac{S}{8}[F_\m Q(\omega)+F^+_\m Q(\omega)-G_\m Q(\omega)-G_\m Q(-\omega)].
\end{equation}
 Functions $I m \chi_\perp$ and $I m C$ determine parts of the neutron scattering cross section independent on the neutron polarization $\m P_0$ and proportion to it respectively (See for example \cite{mal1}). The  chiral contribution to the cross section   appears due to the DM interaction and is $\m q$ odd in agreement with general theory \cite{mal1}.
 
In the EPR case when $\m q\rightarrow 0$ the singular term $\m q^2_\perp/|\m{q, k}|^2$ in Eqs. (20) and  (21) is equal to $N_\perp/2=N_{a a}+N_{b b}$ and $\hat q^2_c=N_{c c} $ respectively. At the same time the umklapp interaction $|R|=S\omega_0(N^2_{ac}+N^2_{bc})^{1/2}/4$ is small  and we can neglect it in the dispersion Eq.(B4).  As a result using Eqs.(20), (21) and (31) we obtain
\begin{equation}
	\chi^{A A^*}_{\alpha\beta}(\omega)=-\frac{1}{2}(\delta_{\alpha\beta}-\hat c_\alpha\hat c_\beta)\frac{S
(2A k^2+S \omega_0 N_{cc})}{\omega^2-\epsilon^2_1}, 
\end{equation}
where 
\begin{equation}
	\epsilon_1=[2(A k^2)^2+S\omega_0 A k^2+(S\omega_0)^2N_{cc} N_\perp/2]^{1/2}.
\end{equation} 
and $C=0$. If we take into account the umklapp interaction the resonances  appear at $\omega=\Delta$ and $\epsilon_{2\m k}$ and so on.

If $\m q\neq 0$ as above we consider two cases: $\m{q\parallel k}$ and $\m{q\perp k}$. In the first case we have
\begin{equation}
	\chi_\perp=-\frac{S}{4}\left(\frac{Z_-}{\omega^2-\epsilon^2_-}+\frac{Z_+}{\omega^2-\epsilon^2_+}\right);\quad C=-\frac{S}{4}\left(\frac{Z_-}{\omega^2-\epsilon^2_-}-\frac{Z_+}{\omega^2-\epsilon^2_+}\right),
\end{equation}
where $Z_\pm=A(k\pm q_\parallel)^2+S\omega_0$ and $\epsilon^2_\pm={A(k \pm q_\parallel)^2[A(k\pm q_\parallel)^2+A k^2+S \omega_0]} $. 

For $\m{q\perp k}$ using (C5) we obtain
\begin{equation}
\chi_\perp=-\frac{S}{4(\epsilon^2_+-\epsilon^2_-)}\left[\frac{Z_+}{\omega^2-\epsilon^2_+}+\frac{Z_-}{\omega^2-\epsilon^2_-}\right],
\end{equation}
where  $\epsilon_\pm$ are given by Eq. (35), $Z_\pm=\pm[2A^2 q^2_\perp(2k^2+q^2_\perp)]\pm(\epsilon^2_\pm-\epsilon^2_1)$ and $C=0$.

We discuss now briefly off-diagonal terms in Eq.(60). They are a result of the umklapp interaction and has to be less than that considered above. At $\m q=0$ the $\hat c \m A$ and $\m{A A}$ terms are proportional to  off-diagonal components $N_{ca(b)}$ of the demagnetization tensor in the first and the second degree respectively.
Possibility of their experimental study is a special problem which is beyond scope of this paper. At $\m q\neq 0$ these terms are very complex and we do not analyse them here too.

\underline{\textit{"Ferromagnetic" state ($H>H_c$)}}
In the longitudinal $cc$ channel there are the two-spin excitations only which are beyond our consideration. By the same way as above for the transverse susceptibility we obtain 
\begin{equation}
	\chi_{\perp \alpha\beta}=-\frac{S}{2}(\delta_{\alpha\beta}-\hat c_\alpha \hat c_\beta)\left(\frac{1}{\omega-\epsilon_\m{q+k}}-\frac{1}{\omega+\epsilon_\m{q-k}}\right)-\frac{i S}{2}\epsilon_{\alpha\beta\gamma} \hat c_\gamma\left(\frac{1}{\omega-\epsilon_\m{q+k}}+\frac{1}{\omega+\epsilon_\m{q-k}}\right),
\end{equation}
 where the second term represents the chiral contribution to the susceptibility and  the spin-wave energy is given by Eq.(22). At $\m q=0$ the last term in (22) contains the demagnetization and $\epsilon _\m k=A k^2+S\omega_0(N_{a a}+N_{b b})/2+g\mu_B(H-H_c\chi)$. In this case  the imaginary part of the chirality is $\omega$-even and the chiral contribution to the static susceptibility is equal to zero in agreement with the general theory \cite{mal1}. Experimentally this contribution could be measured using circularly polarized AC field. For $\m q\neq 0$ the neutron scattering is maximal at $\m{q=\pm k}$ in contrast to conventional ferromagnets.

\section{Summary and Discussion}
We begin with  a short survey of the main results presented above. We have used the following interactions: conventional isotropic exchange, the DMI, anisotropic exchange, magnetic dipole interaction and Zeeman energy. In the classical approximation these interactions determine the helix form and the critical field $H_c$ of the transition to the "ferromagnetic" spin configuration. This field depends on the sample form due the demagnetization.

 The linear spin-wave theory was developed. It was shown that the spin-wave spectrum depends strongly on the magnetic field. At $H>H_c$ we have quadratic spectrum  with the gap linearly increasing with the field [see Eq.(22)]. Below $H_c$ the spectrum is gapless. It is  strongly anisotropic due to incommensurate helical structure and low symmetry of the DMI. As a result umklapp processes appear which   connect the spin-waves  with momenta $\m q$ and $\m{q\pm k}$ and  different energies.  For arbitrary $\m q$  the energy has very complex form. It is determined as a solution of  infinite set of linear equations connecting the states with $\m q$ and $\m q\pm n\m k$ where $n=1,2,...$. Restriction to $n=1$ leads to six equations which general solution remains  complex.  For $\m{q\perp k}$ there are two modes given by Eq.(35). One  has  the gap equal to $A k^2$ where  $A $ is the spin-wave stiffness at  $q\gg k$.  The second mode  is the gapless with quadratic dispersion at small $q$. Both gapless branches (parallel and perpendicular to $\m k$) and the gapped one are shown in Fig.1. At $q\gg k$ all branches merge and the anisotropy of the spectrum disappears. The $n=2$ approximation do not  changes these results qualitatively. For  $q\ll k $ at $H=0$ the spectrum has  simple form  given by Eq.(36).
 
 The classical energy depends on the field component along the helix axis $\m k$ only. However it was shown experimentally that rather weak perpendicular field $H_\perp\ll H_c$ rotates the helix   and its axis is settled  along the field \cite{leb} \cite{ok}. This quantum phenomenon is a consequence  of the spiral  spin structure when the angle between spin and the field depends on the lattice point. As a result we get the umklapps again and the spin-wave spectrum in the gapless case is unstable at infinitesimal perpendicular field. So we must introduce the gap $\Delta$ "by hands"  and the spectrum becomes stable if $g\mu_B H_\perp<\Delta$ [see Eq.(33)].  The magnetization along $\m H_\perp$  appears  and the  helix becomes  deformed (see Eqs.(46) and (47) respectively). The second harmonic with the wave-vector $2\m k$ was observed in \cite{leb} and \cite{ok}.
 
 We considered two ways of the gap origin: cubic anisotropy and interaction between spin waves.  The first contribution is proportional to the strength $K$ of the cubic anisotropy. The second appears is a result of breaking of the total spin conservation law by the DMI. It is positive and disappears in the "ferromagnetic state" at $H>H_c$ due to absence of the zero-point vibrations. But in this region there is the field-induced gap determined by Eq.(22). In ferromagnets the sign of $K$ determines the direction of the easy axis. In our case  the helical structure is stable if $\Delta^2$ given by Eq.(57) is positive. Otherwise we have the chiral spin liquid. So there is  a question : If  change sign of $\Delta^2$ is a reason of the transition to the disordered state in $M n S i$ at high pressure? This problem demands further experimental and theoretical study.

The helix structure leads to peculiar  features  of the ESR and neutron scattering. In conventional magnets the ESR frequency is equal to the spin-wave gap. In the helical systems  due to the umklapps  along  with this frequency there are more higher resonant excitations corresponding to the spin-waves with $\m q=n\m k$ where $n=1,2,...$. At $\m H\neq 0$ the chiral channel appears.  It may be observed using circular radiation. In neutron scattering  below $H_c$ three modes can be studied.  The chiral channel exists at $\m H=0$ and $\m q\neq 0$ due to axial-vector nature of the DMI.

Detailed experimental work was done in the case of $M n S i$ compound only. We now compare some of known experimental results obtained at ambient pressure with the predictions of our theory and discuss possibilities of the further experimental studies. The principal parameters are: lattice spacing $a=4.558\AA$, $T_c\simeq 29K$, $k\simeq 0.035 \AA^{-1}$, saturated magnetization $M=0.4\mu_B/a^3\simeq 0,016T,\quad 4\pi M=0.20T$, critical field $H_c=0.5\div 0.6T$ \cite{koy} and spin-wave stiffness $A\simeq 52meV \AA^2$ \cite{is}. From these data and Eq.(12) we obtain $H_c\simeq A k^2/(g\mu_B)\simeq 0.55T$. This value coincides with experimentally observed  critical field. For more precise comparison one must measure all parameters including the demagnetization $N_{cc}$ using single sample.

To the best of my knowledge the ESR in $M n S i$ was studied  only in \cite{date}. Several resonances were observed but only one was studied qualitatively as a function of the magnetic field. Its  frequency in zero field is equal to $0.93T$. Using Eq.(70) and taking into account that $S\omega_0=4\pi g\mu_B M$  we obtain $0.85T$. The agreement is within the error bars. According to Eq.(22) at $H=H_c$ the frequency is close to $0.6T\simeq H_c$ and then increases linearly with $H$. This behavior coincides with results of Ref. \cite{date}. Further experimental studies are necessary. First of all it is essential to observe the gap  $\Delta $ and its dependence on the perpendicular field, which accordingly with Eq.(33) should be   $\Delta=[\Delta^2-(1/2)(g\mu_B H_\perp)^2\cos^4\alpha]^{1/2}$. The observation of the umklapp resonances $\epsilon_{2\m k}$ and $\epsilon_{3\m k}$ would be important too.

The theory developed in this paper  explains partly some experimental findings. It was developed in the linear spin-wave approximation.  The spin-wave interaction  was used for evaluation contribution to the gap only.  We also did not evaluate some experimentally observed quantities such as specific heat and the site magnetization. Corresponding results will be published elsewhere.

\section{Acknowledgments}   
I am very grateful to A.I. Okorokov,  S.V.Grigoriev and P.B\"{o}ni for numerous interesting discussions of the $M n S i$ problem. The work was supported by RFBR (Grand Nos SS-1671.2003.2, 03-02-17340 and 00-15-96814), Grant Goscontract 40.012.1.1.1149 and Russian Programs "Quantum Macrophysics", "Strongly correlated electrons in semiconductors, metals, superconductors and magnetic materials", and "Neutron research of solids".
   
\appendix
\section{Equations of motion}
Using Eqs.(25), (27) and (28) in the $n=1$ approximation for functions (29) we obtain
\begin{equation}
	\begin{aligned}
	(\omega-E)G-B F-[P c+R_\m{-q}(1-s)-R_\m{q-k}(1+s)]G_- +(R_\m{-q}+R_\m{q-k})(1+s)F_-\\
	-[P^* c+R^*_\m{-q-k}(1-s)-R^*_\m q
(1+s)]G_+-(R^*_\m q +R^*_\m{-q-k})(1-s)F_+&=1,\\
B G+(\omega+E)F+(R_\m{-q}+R_\m{q-k})(1-s)G_-+[P c+R_\m{q-k}(1-s)-R_\m{-q}(1+s)]F_-\\
-(R^*_\m q+R^*_\m{-q-k})(1+s)G_-+[P^* c+R^*_\m q(1-s)-R^*_\m{-q-k}(1+s)]F_+&=0,\\
-[P^* c+R^*_\m{-q}(1-s)-R^*_\m{q-k}(1+s)]G-(R^*_\m{-q}+R^*_\m{q-k})(1-s)F+(\omega-E_-)G_--B_-F_-&=0,\\
-(R^*_\m{-q}+R^*_\m{q-k})(1+s)G+[P^* c+R^*_\m{q-k}(1-s)-R^*_\m{-q}(1+s)]F+B_-G_-+(\omega+E_-)F_-&=0,\\
-[P c+R_\m{-q-k}(1-s)-R_\m q(1+s)]G+(R_\m q+R_\m{-q-k})(1+s)F+(\omega-E_+)G_+-B_+F_+&=0,\\
(R_\m q+R_\m{-q-k})(1-s)G+[Pc+R_\m q(1-s)-R_\m{-q-k}(1+s)]F+B_+G_++(\omega+E_+)F_+&=0,
	\end{aligned}
\end{equation}
where $s=\sin\alpha$, $c=\cos\alpha$, $E(B)=E(B)_\m q$,  $E(B)_\pm=E(B)_\m {q\pm k}$ and functions $E$ and $B$ are determined by Eqs. (18) and (19).

In matrix form these equations are given by
\begin{equation}
	M(\m q, \omega)=I
\end{equation}
 where $I=(1,0,0,0,0,0)$. Determinant of the matrix $M$ is even function of $\omega$ and has the following general form
\begin{equation}
	Det[M]=(\omega^2-\epsilon^2_{\m q 0})(\omega^2-\epsilon^2_{\m q +})(\omega^2-\epsilon^2_{\m q -})
\end{equation}

We  obtain three renormalized energies  as roots of this equation. As $Det[M]$ is the denominator of the Green functions they have three poles.  However at $\m q\perp \m k$  two  initial spin-wave energies are equal and only two renormalized branches are physically relevant. Hence the Green functions must have two poles. That it is the case is demonstrated below in Appendixes B and C. There are five  renormalized branches in the $n=2$ approximation (see Appendix C) and so on. 

 General expressions for $\epsilon^2_{\m q 0}$ and $\epsilon^2_\m{q \pm }$ are  very complex. We consider below two main cases: i. Small $\m q\ll k$  and ii. $H=0$, $\m q\sim \m k$.  
\section{The $\m q\rightarrow 0$ case}
From Eqs. (20) and (21)  we have  
\begin{equation}
	\begin{aligned}\\
	E_\m q&=A q^2+(A k^2\cos^2\alpha)/2+(S\omega_0/2)\hat q_c^2\cos^2\alpha, &B_\m q&=(1/2)(A k^2+S\omega_0 \hat q_c^2)\cos^2\alpha,\\
	E_1&=A k^2(2+\cos^2\alpha)/2+(S\omega_0/2)[\hat q^2_c\cos^2\alpha+(\hat q^2_\perp/2)(1+\sin^2\alpha)], &B_1&=(1/2)[A k^2+(S\omega_0/2)(\hat q^2_c-\hat q^2_\perp/2)]\cos^2\alpha,
	\end{aligned}
\end{equation}
where  $E_1=E_\m{q\pm k}$ and $B_1=B_\m{q\pm k}$. From Eq. (27) we get 
	$R_\m q=(S\omega_0/2)(\hat c\cdot \hat q)(\hat q\cdot\m A)\cos\alpha$ and $R_\m{q\pm k}=0$. If $\m q\equiv 0$ we must replace $\hat q^2_c$, $\hat q^2_\perp$ and $(\hat c\cdot \hat q)(\hat q\cdot \m A) $ by $N_{c c}$,  $N_{a a}+N_{b b}$ and $(N_{c a}-i N_{c b})/2$ respectively. The matrix $M$ has now the form
	\begin{equation}
\begin{pmatrix}
\omega-E_\m q&-B_\m q&-P c-R(1-s)&R(1+s)&-P^* c+R^*(1+s)&-R^*(1-s)\\
B_\m q&\omega+E_\m q&R(1-s)&P c-R(1+s)&-R^*(1+s)&P^*c+R^*(1-s)\\
-P^*c-R^*(1-s)&-R^*(1-s)&\omega-E_1&-B_1&0&0\\
-R^*(1+s)&P^*c-R^*(1+s)&B_1&\omega+E_1&0&0\\
-Pc+R(1+s)&R(1+s)&0&0&\omega-E_1&-B_1\\
R(1-s)&Pc+R(1-s)&0&0&B_1&\omega+E_1\\
\end{pmatrix}
\end{equation}
 For the determinant of this matrix using Program Mathematica 5 we obtain
\begin{equation}
\begin{aligned}
	Det[M]&=(\omega^2-\epsilon^2_1)\{(\omega^2-\epsilon^2_1)(\omega^2-\epsilon^2_\m q)-4|P|^2(E_0 E_1+B_0 B_1-|P|^2\cos^2\alpha+\omega^2)\cos^2\alpha-8|R|^2(E_0-B_0)(E_1+B_1)\\&+4(P R^*+P^*R)[\omega^2+(E_0-B_0)(E_1-B_1)-2|P|^2\cos^2\alpha]\sin\alpha \cos\alpha-8|R|^2 (E_0-B_0)(E_1-B_1)\sin^2\alpha\}+\\&
	4[(P R^*+P^*R)^2\omega^2-4|P|^2|R|^2\epsilon^2_1]\sin^2\alpha \cos^2\alpha,
	\end{aligned}
\end{equation}
 where $\epsilon^2_0=E^2_0-B^2_0\rightarrow \Delta^2 $. Here the last term  may be represented as $\{|P^2||R|^2(\omega^2-\epsilon^2_1)+[(N_{c\perp}\cdot \m P_\perp)^2-N^2_{c\perp}\m P^2_\perp]\omega^2\}\sin^2\alpha \cos^2\alpha$. The $\omega^2$ term breaks rotational invariance in the $ab$ plane and $\epsilon^2_1$ is not a  root of this equation. But this term is very small and we neglect it.    
 
 We assume that  $P$, $R$ and $\Delta$ are small in comparison with $A k^2$ which is main parameter of the  theory. In this approximation we have
\begin{equation}
	Det[M]=(\omega^2-\epsilon^2_1)\left[(\omega^2-\epsilon^2_1)(\omega^2-\Delta^2)-4| P|^2(E_0E_1+B_0B_1+\omega^2)\cos^2\alpha-8|R|^2\frac{E_1(1+\sin^2\alpha)+B_1\cos^2\alpha}{E_0+B_0}\right]
\end{equation}. As complete expressions for resonant modes and the Green functions in this approximation remain very complex we consider two limiting cases: $R=0$ and $\m H=0$.

\textit{\underline{The $R=0$ case}.} As we see below the Green  functions do not contain pole at $\omega^2=\epsilon ^2_1$. For two other energies we have
\begin{equation} 
\begin{aligned}
	\epsilon^2_0&=\Delta^2-\frac{h^2_\perp(A k^2+S\omega_0 N_{cc})\cos^4\alpha}{2(A k^2+S\omega_0 N_\perp/2\cos^2\alpha)}\simeq \Delta^2-\frac{h^2_\perp\cos^4\alpha}{2},\\&
	\epsilon^2_+=\epsilon^2_1+h^2_\perp\left[1+\frac{(A k^2+S\omega_0 N_{cc})\cos^2\alpha}{2A k^2+S\omega_0 N_\perp \cos^2\alpha}\right]\cos^2\alpha\simeq \epsilon^2_1+\frac{h^2_\perp(2+\cos^2\alpha)}{2}\cos^2\alpha,
	\end{aligned}
\end{equation}
where $\epsilon_1$ is given by Eq.(67) and in right hand side we neglected the dipole contribution.
	
	Corresponding expressions for the Green functions are given by
\begin{equation}
	\begin{aligned}
	G(\omega)&=Z^{-1}[(\omega+E_0)(\omega^2-\epsilon^2_1)+(h^2_\perp/2)(E_1-\omega)],\\
	F(\omega)&=-Z^{-1}B_0[\omega^2-\epsilon^2_1-(h^2_\perp/2)B_1\cos^2\alpha],\\
	G_-(\omega)&=Z^{-1}(h_+/2)[\omega^2-\epsilon^2_1+E_0E_1+B_0B_1-(h^2_\perp/2)\cos^2\alpha]\cos\alpha,\\
	F_-(\omega)&=-Z^{-1}(h_+/2)[B_0E_1+B_1E_0+\omega(B_1-B_0)]\cos\alpha,
	\end{aligned}
\end{equation}
where $Z=(\omega^2-\epsilon^2_+)(\omega^2-\epsilon^2_0)$, $ h_\pm=(h_a\pm i h_b)/2$ and one gets $G_+$ and $F_+$ replacing $ h_+$ by $h_-$ in the expressions for $G_-$ and $F_-$. Neglecting the dipole interaction we obtain (32)

Solution of Eq.(39) for the spin deviations frozen in the perpendicular field is given by
\begin{equation}
	\begin{aligned}
	a_\m k&=(a^+_\m k)^*=-\sqrt{\frac{S}{2}}\frac{(\m A\cdot\m h)(1+\cos^2\alpha+\sin\alpha)\Delta^2}{A k^2(1+\cos^2\alpha)[\Delta^2-(h^2_\perp/2)\cos^4\alpha]},\\&
	a_\m{-k}=(a^+_\m{-k})^*=\sqrt{\frac{S}{2}}\frac{(\m A^*\cdot\m h)(1+\cos^2\alpha-\sin\alpha)\Delta^2}{A k^2(1+\cos^2\alpha)[\Delta^2-(h^2_\perp/2)\cos^4\alpha]},\\&
	a_0=a^+_0=\sqrt{\frac{S}{2}}\frac{h^2_\perp(E_0-B_0)\sin\alpha \cos\alpha}{2A k^2[\Delta^2-(h^2_\perp/2)\cos^4\alpha]}.\\
	\end{aligned}
\end{equation}

\textit{\underline{The $\m H=0$} case}. From Eq.(B4) we have now
\begin{equation}
	Det[M]=(\omega^2-\epsilon^2_1)(\omega^2-\epsilon^2_0)(\omega^2-\epsilon^2_+),
\end{equation}
where 
\begin{equation}
	\begin{aligned}
	\epsilon^2_0&=\Delta^2\left[1-\frac{(S\omega_0)^2(N^2_{c a}+N^2_{c b})}{(A k^2+S\omega_0 N_{c c})(2A k^2+S\omega_0 N_\perp)}\right],\\&
	\epsilon^2_+=\epsilon^2_1+\frac{\Delta^2(S\omega_0)^2(N^2_{c a}+N^2_{c b})}{(A k^2+S\omega_0 N_{cc})(2A k^2+S\omega_0 N_\perp)}.\\
	\end{aligned}
\end{equation}
Corresponding expressions for the Green functions are given by
\begin{equation}
	\begin{aligned}
G(\omega)&=Z^{-1}[(\omega+E_0)(\omega^2-\epsilon^2_1)+4|R|^2(E_1+B_1)],\\
F(\omega)&=-Z^{-1}[B_0(\omega^2-\epsilon^2_1)+4|R|^2(E_1+B_1)],\\
G_-(\omega)&=R^*(\omega+E_0-B_0)(\omega+E_1+B_1),\\
F_-(\omega)&=-R^*(\omega+E_0-B_0)(E_1+B_1-\omega),	
	\end{aligned}
\end{equation}
where $R=(S\omega_0/4)(N_{c a}-i N_{c b})$, $G_+=-(R/R^*)G_-$ and $F_+=-(R/R^*)F_-$. Note that now it is  addition sign minus in expressions for $G_+$ and $F_+$ in comparison with the $R=0$ case.

\section{The $q\gtrsim k$ case}.
It is convenient now to use dimensionless variables $X=\omega/(A k^2)$, $Y=|q_\perp|/k$, $V=q_\parallel/k$ and $U=-(q_a-iq_b)/k$. In these variables we have $M=(A k^2)^6 m $ and for $m$ we get
\begin{equation}
	\begin{pmatrix}
	X-W^2-\frac{1}{2}&-\frac{1}{2}&U&0&U^*&0\\
	\frac{1}{2}&X+W^2+\frac{1}{2}&0&U&0&U^*\\
	U^*&0&X-W^2+2V-\frac{3}{2}&-\frac{1}{2}&0&0\\
	0&U^*&\frac{1}{2}&X+W^2-2V+\frac{3}{2}&0&0\\
	U&0&0&0&X-W^2-2V-\frac{3}{2}&-\frac{1}{2}\\
	0&U&0&0&\frac{1}{2}&X+W^2+2V+\frac{3}{2},\\
		\end{pmatrix}
\end{equation}
where $W^2=Y^2+V^2$.\\
 General expression for  $Det[m]$ is very complex. For $V\ll 1$  we have
\begin{equation}
	\begin{aligned}
	Det[m]&=(X^2-\epsilon^2_1)(X^2-\epsilon^2_+)(X^2-\epsilon^2_-)\\
	\epsilon^2_\pm&=1+4Y^2+Y^4\pm( 1+8Y^2+17Y^4+8Y^6)^{1/2}+4V^2\\
	\end{aligned}
\end{equation}
where $\epsilon^2_1=(Y^2+1)(Y^2+2)$ is the energy at $\m {q_\perp\pm k}$ in the dimensionless units. As above at $Z=0$ the factor $X^2-\epsilon^2_1$cancels in the expressions for Green Functions.

 For small  $Y$ and $V$  we have 
\begin{equation}
X^2_-=\frac{Y^4}{2}+V^2;\quad X^2_+=2+4Y^2.
\end{equation}
Asymptotic expressions  for $\epsilon_\pm$ at $Y\gg 1$ are given by
\begin{equation}
\epsilon_\pm=Y^2\pm \sqrt{2}Y.	
\end{equation}

The Green functions in dimensionless units are given by
\begin{equation}
\begin{aligned}
	G(X)&=Z^{-1}[(X+Y^2+1/2)(X^2-\epsilon^2_1)+2Y^2(Y^2+3/2-X)];\\
	F(X)&=-Z^{-1}(X^2-\epsilon^2_1+2Y^2)/2;\\
	G_-(X)&=-Z^{-1}(Y_a+i Y_b)[-(X+Y^2+1/2)(X+Y^2+3/2)+2Y^2+1/4];\\
	F_-()&=Z^{-1}(Y_a+i Y_b)(2X-1)/2,
	\end{aligned}
\end{equation}
 where $\epsilon^2_1=(Y^2+1)(Y^2+2)$, $Z=(X^2-\epsilon^2_-)(X^2-\epsilon^2_+)$  and $G_+(F_+)=[G_-(F_-)](Y_a-i Y_b)/(Y_a+i Y_b)$.
 
To illustrate that $n=1$ results are at least qualitatively correct we consider now the $n=2$ approximation. In this case the matrix $M=(A k^2)^{10}m$ and for $m$ we have
\begin{equation}
	\begin{pmatrix}
	X_--\frac{1}{2}&-\frac{1}{2}&U&0&U^*&0&0&0&0&0\\
	\frac{1}{2}&X_++\frac{1}{2}&0&U&0&U^*&0&0&0&0\\
	U^*&0&X_--\frac{3}{2}&-\frac{1}{2}&0&0&U&0&0&0\\
	0&U^*&\frac{1}{2}&X_++\frac{3}{2}&0&0&0&U&0&0\\
	U&0&0&0&X_--\frac{3}{2}&-\frac{1}{2}&0&0&U^*&0\\
	0&U&0&0&\frac{1}{2}&X_++\frac{3}{2}&0&0&0&U^*\\
	0&0&U^*&0&0&0&X_--\frac{9}{2}&-\frac{1}{2}&0&0\\
	0&0&0&U^*&0&0&\frac{1}{2}&X_++\frac{9}{2}&0&0\\
	0&0&0&0&U&0&0&0&X_--\frac{9}{2}&-\frac{1}{2}\\
	0&0&0&0&0&U&0&0&\frac{1}{2}&X_++\frac{9}{2}\\
	\end{pmatrix}
\end{equation}
 where $X_\mp=X\mp Y^2$. The determinant of  $m$ is given by
\begin{equation}
	\begin{aligned}
	Det[m]&=1600X^2-1760 X^4+564 X^6-44 X^8+X^{10}+(7760X^2-4868X^4+524X^6-15X^8)Y^2\\&
+(-600+9279X^2-2462X^4+167X^6-3X^8)Y^4+(-585+3686X^2-597X^4+22X^6)Y^6\\&+(-519+866X^2-107X^4+3X^6)Y^8+(-528+128X^2-7X^4)Y^{10}+(-144+24X^2-X^4)Y^{12}.
	\end{aligned}
\end{equation}
We see that according to argumentation in Sec.5  we have cancellation of the $Y^2$ term. There is the $Y^2X^2$ term only.

The equation $Det[m]=0$ determines now five spin-wave energies generated  by  $\epsilon^2_\m q=Y^2(Y^2+1)$,  $\epsilon^2_\m k=(Y^2+1)(Y^2+2)$ and $\epsilon_\m{2k}=(Y^2+4)(Y^2+5)$. We consider below the three lower branches only.

\underline{\textit{Small $Y$case}}. From Eq.(C6) for the gapless branch we have $X^2=3Y^4/8$ and the factor $1/2$ in (C3) has to be replaced by $3/8$. Higher approximations with $n>2$ can not change this result. Consider now the  branches connected to $\epsilon_\m k$. For  $X^2\simeq 2$  the first three terms in Eq.(C7)  give  $648\chi^2-5904\chi Y^2+9398Y^4= 0$, where $\chi=X^2-2$. Solutions of this equation are given by
\begin{equation}
	X^2=2+\frac{37Y^2}{18}=2+2.1Y^2;\quad X^2=\frac{127Y^2}{18}=2+7.1Y^2.
\end{equation}
We again obtain small changes of the numerical coefficients.

\underline{\textit{Large $Y$}}.
For large $Y$ we can write $X=W Y^2$ and asymptotically $W\rightarrow 1$. Retaining only three main terms  in  powers of $Y$ we obtain the dispersion equation
\begin{equation}
	(W-1)^3-\frac{8(W-1)+15(W-1)^2}{Y^2}+\frac{40}{Y^4 W^2}=0.
\end{equation}
Its solutions are given by
\begin{equation}
	X_\pm=Y^2\pm Y\sqrt{2};\quad X_1=Y^2+\frac{5}{2}.
\end{equation}

\section{The Hartree-Fock gap}
As the interaction (55) is non-Hermitian the spin-wave Hamiltonian  have the form \cite{pet}
\begin{equation}
	H_{SW}=\sum[E_\m q a^+_\m q a_\m q+\frac{1}{2}(B^+_\m q a_\m q a_\m{-q}+B_\m q a^+_\m{-q}a^+_\m q)],
\end{equation}
	where $B^*_\m q\neq B^+_\m q$. Corresponding corrections to $E_\m q$ and $B_\m q$ in Eq. (19) are determined as
\begin{equation}
	\delta E_\m q=\left<\frac{\delta^2 V_I}{\delta a^+_\m q \delta a_\m q}\right>;\quad \delta B_\m q=\left<\frac{\delta^2 V_I}{\delta a^+_\m q \delta a^+_\m{-q}}\right>;\quad\delta B^+_\m q=\left<\frac{\delta^2 V_I}{\delta a^+_\m{-q}\delta a^+_\m q}\right>.
\end{equation}
At $\m q=0$ we obtain
\begin{equation}
\begin{aligned}
	\delta E_0&=-\frac{1}{2}\sum[(V_0+V_\m q-J_\m q+J_0)n_\m q+(\frac{1}{2}V_0+V_\m q-J_\m q+J_0)f_\m q];\\&
	 \delta B_0=-\frac{1}{2}\sum V_\m q f_\m q;\quad \delta B^+_0=\delta B_0-\frac{1}{2}\sum[(V_0+V_\m q-2J_\m q+2J_0)n_\m q-(J_\m q-J_0)f_\m q],
	 \end{aligned}
\end{equation}
where expression for $V_\m q$ is given below Eq.(55) and
\begin{equation}
 n_\m q=\frac{E_\m q-\epsilon_\m q}{2\epsilon_\m q};\quad	f_\m q=<a_\m q a_\m{-q}>=<a^+_\m{-q}a^+_\m q>=-\frac{B_\m q}{2\epsilon_\m q}
\end{equation}
Using Eqs. (20) and (21) and neglecting dipolar terms we get
\begin{equation}
	\Delta^2_I=-\frac{A k^2}{4}\sum[(V_0+V_\m q)n_\m q+(V_0-J_\m q+J_0)f_\m q].
	\end{equation}
	Large $q\sim 1/a$ give principal contribution to the expression (D 3). However  Eqs. (20) and (21) determine $E_\m q$ and $B_\m q $ at $q\ll 1/a$ only. Using Eq. (55) we obtain these functions for all $q$
\begin{equation}
	E_\m q=S\left[J_0-\frac{J_\m{q,k}+J_\m q}{2}+(D_0-D_\m q)\frac{A k^2}{2}\right];\quad B_\m q=\frac{S}{2}\left[J_\m{q,k}-J_\m q+2\frac{D_\m q}{S D_0}A k^2\right],
\end{equation}
 where using Eq.(8) we replaced $\m{k\cdot[\hat a\times \hat b]}$ by $A k^2/(S D_0)$. \\
 The function $B_\m q\propto k^2$. As $\epsilon_\m q=(E^2_\m q-B^2_\m q)^{1/2}$  we have that $n_\m q$
 is proportional to $(A k^2)^3$ and may be neglected. In the $f_\m q$ term at large $q$ we have $V_0+J_0-J_\m q\simeq \epsilon_\m q/S$ and
\begin{equation}
	\Delta^2_I=\frac{A k^2}{8}\sum\left[J_\m {q,k}-J_\m q+\frac{2D_\m q A k^2}{S D_0}\right],
\end{equation}
 where the last term contribute to the gap only as $\sum J_\m{q,k}=0$.

\begin{figure}
\centering
\includegraphics[scale=0.7]{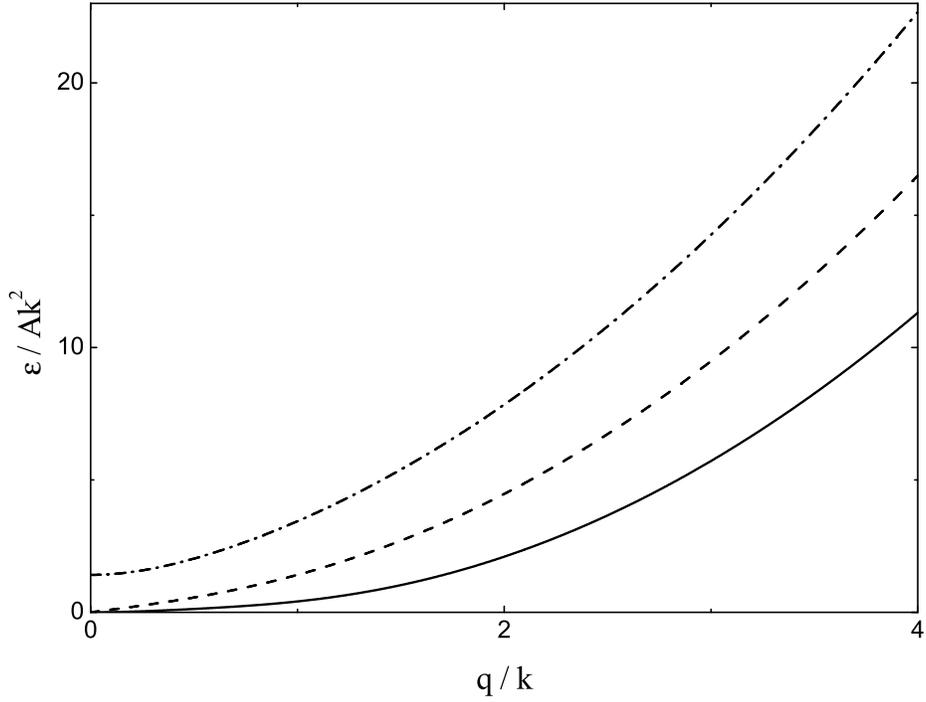}
\caption{Fig.1 Spin-wave dispersion for different directions of the wave-vector $\m q$. (a) $\m q$ is along the helix axis $\m k$ (dashed line). (b) Gapless and gapped branches for $\m{q\perp k}$. Solid and dot-dashed lines respectively.}
\end{figure}
  

\begin{thebibliography}{99}
\bibitem{demag} At $\m q=0$  we have  $\hat q_\alpha\hat q_\beta\rightarrow N_{\alpha\beta}$ where $N_{\alpha\beta}$ is the tensor of demagnetization.
\bibitem{def} Conventionally normalized magnetization is equal to $g\mu_B\m S^\perp_U/a^3$ where  $a$ is the lattice constant.
\end{thebibliography}
\end{document}